\documentclass[letterpaper, 10pt]{IEEEtran}

\usepackage{subcaption}
\usepackage{url}
\usepackage{booktabs}
\usepackage{graphicx}
\usepackage{xcolor}
\begin{document}

\title{Analysis of Location Data Leakage in the Internet Traffic of Android-based Mobile Devices}

\author{\IEEEauthorblockN{Nir Sivan, Ron Bitton, Asaf Shabtai\\}
\IEEEauthorblockA{Department of Software and Information Systems Engineering\\
Ben-Gurion University of the Negev\\
Beer-Sheva, 8410501, Israel\\}

\{sivanni,ronbit\}@post.bgu.ac.il, shabtaia@bgu.ac.il
}

\maketitle

\begin{abstract}
In recent years we have witnessed a shift towards personalized, context-based applications and services for mobile device users.
A key component of many of these services is the ability to infer the current location and predict the future location of users based on location sensors embedded in the devices.
Such knowledge enables service providers to present relevant and timely offers to their users and better manage traffic congestion control, thus increasing customer satisfaction and engagement.
However, such services suffer from location data leakage which has become one of today's most concerning privacy issues for smartphone users.
In this paper we focus specifically on location data that is exposed by Android applications via Internet network traffic in plaintext (i.e., without encryption) without the user's awareness.
We present an empirical evaluation, involving the network traffic of real mobile device users, aimed at: (1) measuring the extent of location data leakage in the Internet traffic of Android-based smartphone devices; and (2) understanding the value of this data by  inferring users' points of interests (POIs).
This was achieved by analyzing the Internet traffic recorded from the smartphones of a group of 71 participants for an average period of 37 days.
We also propose a procedure for mining and filtering location data from raw network traffic and utilize geolocation clustering methods to infer users' POIs.
The key findings of this research center on the extent of this phenomenon in terms of both ubiquity and severity; we found that over 85\% of devices of users are leaking location data, and the exposure rate of users' POIs, derived from the relatively sparse leakage indicators, is around 61\%.
\end{abstract}

\section{\label{sec:introduction}Introduction}
In recent years, there has been a trend towards the personalization of services in many areas.
This is particularly true for services provided on mobile devices, where a plethora of context-based applications (for example, Yelp, Uber, Google Maps, and Google Now) are used daily by millions of people.
These devices possess a tremendous amount of private information, ranging from users' personal and financial data to their location data, making such devices the target of commercial entities and intelligence gathering.
A key property of many of these services is the ability to understand the users' current location, infer points of interest, and predict the future location of users based on location sensors embedded in the devices.
Such knowledge enables service providers to present relevant and timely services (such as navigation recommendations, weather forecasts, advertisements, social networks, etc.) to their users, thus increasing customer satisfaction and engagement.
Methods for deriving the location of a mobile device can be categorized into the following two approaches:

\begin{itemize}

\item \textbf{Host-based:}
In this approach, an installed application can infer the location of the device by probing built-in sensors or evaluating data provided when a user checks in to a place on social media.
Local sensors that can provide location data include hotspot (Wi-Fi) information such as the SSID
and BSSID \cite{Nguyen:2013}, connected cellular, as well as GPS \cite{hazan2017dynamic}.
The location can also be inferred by using various side-channel attacks such as power supply variance analysis \cite{michalevsky_powerspy:_2015}.

\item \textbf{Network-based:}
In this approach, the location of the mobile device can be derived by using cell tower triangulation \cite{arigela_mobile_2013} (i.e., using radio-location by analyzing signals received by the cell towers the device is connected to).
This requires high privileged access to the data, which is usually available to service providers and law enforcement agencies.

\end{itemize}

In order to utilize location traces as a meaningful information source, it is imperative to analyze the data and aggregate it to location clusters that are important to the user, such as home, shopping, or work \cite{hazan2017dynamic}.
These locations are also known as the users' points of interest (POIs).
The most common approach for inferring a user's POIs from location traces is by clustering the location traces by distance and time thresholds; eventually, a cluster will be produced if the user has stayed in the same place for a sufficient amount of time.
POIs are identified by understanding which clusters are important to the user and omitting less important data such as transit data \cite{kang_extracting_2005}\cite{umair_discovering_2014}.

Location data collected on the host (mobile device) may be provided to third party services by applications or leaked by a malicious application \cite{ren_recon:_2016}. 
Recent works have reported a high rate of personal data leakage by popular applications over insecure communication channels without users' awareness.
These studies also showed that location data is one of the most "popular" leaked personally identifiable information (PII), as 10\% of the most popular applications leak location data in plain text \cite{ren_recon:_2016}.
In fact, according to Trend Micro, the location permission was identified as the most abused Android application permission.\footnote{\url{http://about-threats.trendmicro.com/us/library/image-gallery/12-most-abused-android-app-permissions}}
This privacy breach was also acknowledged during the 2018 DEFCON workshops, when applications of both iOS and Android-based devices were detected sending accurate location data in unencrypted formats.\footnote{\url{https://arstechnica.com/information-technology/2018/09/dozens-of-ios-apps-surreptitiously-share-user-location-data-with-tracking-firms/}} 

In this research, we investigate the phenomenon of location data leakage in the Internet traffic of Android based smartphones. 
The main goals of this research are as follows:

\begin{enumerate}

\item Understanding the amount and quality of location leakage detected in plain text in the device's network traffic.

\item Analyzing the location leaks in order to infer the user's POIs.
This task is not trivial due to the fact that the vast majority of the previously proposed POI detection methods assume a consistent and high rate of location sampling (e.g., GPS);
therefore, they cannot be directly applied on noisy and sparse location data, such as the data we focus on in this study (i.e., location data leaked over mobile device network traffic).

\item Understanding the privacy exposure level of users in terms of the number of identified POIs, amount of data required for identifying the POIs, accuracy of the detected POIs, and time spent in the POIs.

\end{enumerate}

In order to achieve these goals, we collected and analyzed the Internet traffic of 71 smartphone users for an average of 37 days, while the devices were being used routinely.
In addition, we collected the location of the mobile devices by using a dedicated Android agent (application) that was installed on the devices and sampled the location sensor.
The data collected by the agent was used as the ground truth for the actual location of the mobile device.


In summary the contributions of this paper are as follows:

\begin{enumerate}

\item we explore and discuss the scope, volume and quality of location-based data leaked via insecure, unencrypted network traffic (i.e., as plaintext) of smart mobile devices.

\item we conduct an empirical evaluation based on real data from mobile devices.

\item \textbf{the evaluation involves a unique dataset that was collected simultaneously from the device itself and the network traffic sent from the device; such a dataset from real users' devices is very difficult to obtain.}\footnote{we currently in the process of getting the required permissions for making the anonymized dataset publicly available for research.}

\item we present a methodological process for collecting, processing, and filtering location-based data from mobile devices in order to infer the users' POIs.

\item to the best of our knowledge, we are the first to use POI clustering on a sparse, inconsistent data stream by modifying available clustering algorithms, and discuss the experiments results and the effectiveness of the applied process.
\end{enumerate}

The rest of this paper is organized as follows. 
In Section \ref{sec:relatedworks} we summarize previous related work in the domains of data leakage in mobile devices and inferring points of interest from location traces.
In Section \ref{sec:threatmodel} we describe the assumed threat model, and in Section \ref{sec:framework} we present the data collection framework and the data collection process used in this research, as well as privacy and ethical considerations.
Section \ref{sec:extracting} contains our analysis of the location traces extracted from the captured raw network traffic and Section \ref{sec:infferring} includes an analysis of inferred points of interest from the leaked location traces.
In Section \ref{sec:discussion} we discuss the challenge of identifying the leaking application and possible mitigation strategies, and finally, in Section \ref{sec:futurework} we conclude with a discussion of the paper's contributions and introduce possible directions for future work.

\section{\label{sec:relatedworks}Related Work}

\subsection{Data leakage in mobile devices}

According to the GDPR definition \cite{regulation2016general}, personal data or personally identifiable information (PII) is any data that can be used to identify a person, including name, ID, social media identity and location.
PII leaks are a major privacy concern for mobile devices users.
Along with device and user identifiers, location leakage is among the private data most commonly and extensively leaked from mobile devices \cite{ren_recon:_2016}.
Furthermore, location permission is a very popular permission requested by most mobile apps (25\% of apps use precise location, and an even greater number of apps use coarse location) \cite{taylor_longitudinal_2016}.

Without accurate knowledge about how each app uses and handles its location permissions, access of apps to location API pose a real threat to users' privacy.

There are several ways to abuse network traffic access in order to disclose users' personal information.
For example, one attack model is based on monitoring the network flow behavior and uses machine learning methods to reveal user actions, an approach that is effective even if the traffic is encrypted \cite{grolman_transfer_2018}.
Prior research has also shown that by sniffing the network traffic of well-known and heavily downloaded apps, an attacker can obtain (leak) a broad spectrum of personal and device identifying information without the user's awareness \cite{hodges_android_2014}.
The protocol used to transfer PIIs may be vulnerable to cyber attacks and thus can also be a privacy breach regardless of user awareness.
As a case in point, Ren \textit{et al.} \cite{ren_bug_2018} presenedt a review on data leakage in mobile apps by an unsecured HTTP protocol which can be used to identify the user.

\subsection{Inferring meaningful locations}
Inferring meaningful locations from aggregated location traces is a field of research that has rapidly evolved since the appearance of cheap mobile GPS devices for civilian use.
These devices, as well as smartphones (in the proper navigation and sampling mode) which have become ubiquitous, have a high and constant sampling rate; the availability of a constant data stream is a common assumption for most studies in location analytics.
Inferring meaningful locations relies upon several major algorithm families:
Ester \textit{et al.} introduced DBSCAN \cite{Ester:1996:DAD:3001460.3001507}, which is a density-based algorithm for spatial data, which provides the ability to determine clusters with undefined shapes and is not bound to a specific number of clusters, and does not use temporal data as a parameter. 
Birant \textit{et al.} extended DBSCAN into ST-DBSCAN \cite{birant_st-dbscan:_2007}, which not only uses the spatial data of the database points but also uses its temporal data which is more suitable for spatio-temporal data sets.
Another approach introduced by Kang \textit{et al.} \cite{kang_extracting_2005} clusters places based on time and distance thresholds to determine stay points from transit to improve analysis of trajectories; in the current research we refer to this method as the "incremental method."
Kang \textit{et al.} \cite{kang_extracting_2005} used predefined constraints in order to prevent incorrect clustering due to missing information between traces.
Montoliu \textit{et al.} \cite{montoliu_discovering_2010} also deal with inconsistency and missing data by adding maximum time constraints between traces' constraints.
Alvares \textit{et al.}  \cite{alvares_model_2007} proposed the use of semantic data to better understand the meaning of the collected data, and their method can be used to determine whether a place could be important to the user.       

\section{\label{sec:threatmodel}Threat Model}
Previous research has discussed personal information disclosure and personal information inference from network traffic leakage.
However, those studies mainly dealt with inferring static information, such as demographic attributes or other PII which can be observed when the user is connected to a single malicious hotspot.
In our case, since we are analyzing location data over time in order to obtain contextual information, capturing network traffic from a single hotspot is insufficient.
Thus, in this paper, it is assumed that a threat actor can continuously
capture the network traffic of the user's mobile device.
This special capability is granted to the following main three threat actors:
\begin{enumerate}

\item \textbf{Internet service providers and mobile network operators (MNOs)} which are exposed to the vast amount of the users' network traffic.
The ISP threat model is strong, however we believe that location data leakage is a significant problem by itself and therefore should be explored.
In addition, although ISP's can derive location information from the cell ID, it is very coarse location. 
On the other hand, the location that can be derived from the leaked data is much more accurate.

\item \textbf{VPN and proxy servers}, which relay mobile device network traffic to a third party server and can also misuse the unencrypted data sent in the network traffic.
These solutions are increasingly being used by mobile users for protecting their privacy or consume restricted entertainment content~\cite{khan2018empirical}.\footnote{https://www.forbes.com/sites/forbestechcouncil/2018/07/10/the-future-of-the-vpn-market/}~\footnote{https://blog.globalwebindex.com/chart-of-the-day/vpn-usage/} 

\item \textbf{Tor-like solution}s, which commonly used for protecting privacy~\cite{mani2018understanding}.
In this case the location leakage data can be used by the exit node to expose the real user location (and maybe the user identity).

\end{enumerate}

The main goal of this study is to estimate the potential privacy exposure of a user by the such entities if they choose to misuse this data, or by an attacker.

\section{\label{sec:framework}Data Collection}
In the interest of exploring the extent of location data leakage in the Internet traffic of Android-based smartphones, we developed a dedicated data collection framework.
Using the framework, we collected data from 71 participants for an average period of 37 days.

\subsection{\label{sec:datacollection}Data collection framework}
The framework consists of three main components: a VPN client that collects all of the network traffic transmitted by the device to the Internet, a dedicated Android agent application that obtains location readings from the device's location API, and a light-weight server. 

\begin{itemize}
\item \textbf{VPN client.} The high volume of Internet traffic on smartphones makes it prohibitive to store this information locally on the device for a long period of time.
Alternatively, caching the Internet traffic locally and transmitting it daily to a remote server causes overheads to the device's battery, CPU, and network.
In addition, due to security concerns, such operations require super-user privileges, which necessitate rooting the user's device. 
For these reasons, we opted to use VPN tunneling to redirect the traffic through a dedicated VPN server, where we could record and store the traffic.
It should be mentioned that a VPN is the only user space API provided by the Android operating system to intercept network traffic that does not require super-user or system privileges.

\item \textbf{Android agent (client) application.} To understand the quality of location leakage detected in plaintext within the device's network traffic and to estimate the privacy exposure level of users, we developed a dedicated Android application that obtains the actual location of users.
The application uses the network location provider API which assesses the location by utilizing three sources of information: GPS, Wi-Fi, and Cell ID (obtained from the cellular network).
The data collected by the agent application was used as the ground truth for the actual location of the users.
Applying clustering algorithms on agent-based, accurate and consistent location samples was shown to be accurate and effective for deriving users' POIs. 
Therefore, we opt to use this approach as our baseline and not to rely on the (subjective) collaboration of the participants in providing the actual/labeled POIs.

\item \textbf{Light weight server.} This server has two primary objectives.
First, it operates as an application server, which communicates with the Android agent application and stores all location data in a database.
Second, it acts as a VPN server which communicates with the VPN clients, records all of their network traffic, and redirects it to the Internet.
To provide VPN connectivity and record the traffic, we created a dedicated LAN (local area network) on that server where every VPN client was assigned to a different IP address in the LAN.
The Internet traffic was recorded using "tshark," - a network analyzer tool that is capable of capturing packet data from real-time network traffic.
\end{itemize}

\subsection{Experiment}
In order to achieve the research objectives, we conducted an experiment involving 71 participants.
The participants were current and former students at two universities located in two different cities in Israel. 
Additional information on the participants:
\begin{itemize}

\item 60\% are male and 40\% female,
\item 44\% are in the age of 18 to 24 and 56\% in the age of 25 to 30,
\item 51\% are undergraduate students and 49\% graduate student.

\end{itemize}

\noindent The participants were required to install the two client applications on their personal mobile device for an average period of 37 days.

To ensure that the location sampling had minimal impact on the battery, throughout most of the experiment we set the Android client to sample the location provider for 60 seconds every 20 minutes. In addition, the server was deployed on an AWS (Amazon Web Services) EC2 instance to ensure high availability.

\begin{figure}[tb]
\centering
\includegraphics[scale=0.4]{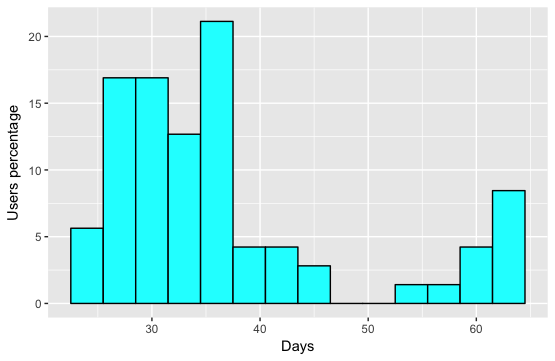}
\caption{The amount of time users participated in the experiment.}
\label{fig:days}
\end{figure}

\begin{figure}[tb]
\centering
\includegraphics[scale=0.4]{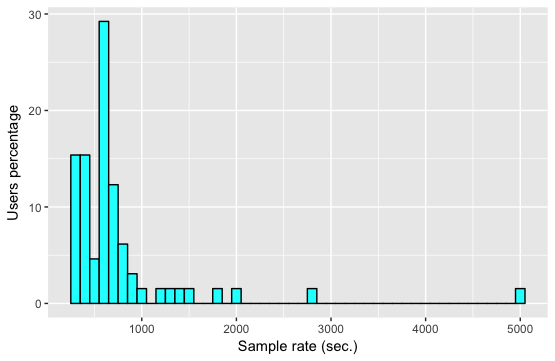}
\caption{The users' actual (averaged) location sample rate of the agent installed on the users' devices.}
\label{fig:rate}
\end{figure}

In Figure \ref{fig:days} we present the amount of time the users participated in the experiment.
In Figure \ref{fig:rate} we present the distribution of the actual location sample rate of the agent application, as observed in the experiment.

Note that the framework that we developed, and specifically the VPN and monitoring agent applications that were installed on the participants smartphones, were used for data collection, research and validation only and are not assumed to be part of the threat model.

\subsection{Privacy and ethical considerations}
The experiment involved collecting sensitive information from real subjects for a long period of time.
To preserve the subjects' privacy, we took the following steps:

\begin{enumerate}

\item The subjects participated in the experiments at their own will and provided their formal consent to participate in the research;
in addition, they were fully aware of the type of data that would be collected and were allowed to withdraw from the study at any time. 
It should be noted that the subjects received a one-time payment as compensation for their participation.

\item Anonymization was applied to the data. At the beginning of the experiment, a random user ID was assigned to each subject, and this user ID served as the identifier of the subject, rather than his or her actual identifying information. 
The mapping between the experiment's user ID and the identity of the subjects was stored in a hard copy document kept in a safe box; at the end of the experiment we destroyed this document.

\item During the experiment, the communication between the agents and servers was fully encrypted. 
In addition, the data collected was stored in an encrypted database.
At the end of the experiment the data was transferred to a local server (i.e., within the institutional network), which is not connected to the Internet. 
Only anonymized information of the subjects was kept for further analysis.

\end{enumerate}

\noindent Based on these steps, the research was approved by the institutional review board (IRB).

\section{\label{sec:extracting}Extracting Location Traces from Raw Network Traffic}
Location data can be transmitted over network traffic in many formats including: explicit geolocation coordinates, names of cities or points of interest, Wi-Fi networks (BSSID), and cellular network data (Cell ID).
In this research we focus on explicit geolocation data that is transmitted in plaintext; such structured data can (potentially) provide more accurate location and is easier to extract and analyze, and thus introduces a greater risk to the user's privacy if leaked. 

\subsection{Process description}
In order to automatically detect location traces within the network traffic of a mobile device, we capture the data at the IP network layer.
Geographic coordinates can be represented in different formats \cite{ISO6709}.
We perform regex search of the standard Android API \cite{Location} representation of geographic coordinates which is decimal degrees in the following format: XX.YYYYYYY.
We specifically used this regular expression for two main reasons. 
First, this is the standard format of the Android location API.
Second, our manual exploration of other location data formats (e.g., names of cities, points of interest, and Wi-Fi networks) indicated that they can dramatically increase the number of false positives (for example, city names sent by a weather forecast application do not indicate the true location of the user).

\begin{figure}[tb]
\centering
\includegraphics[scale=0.97]{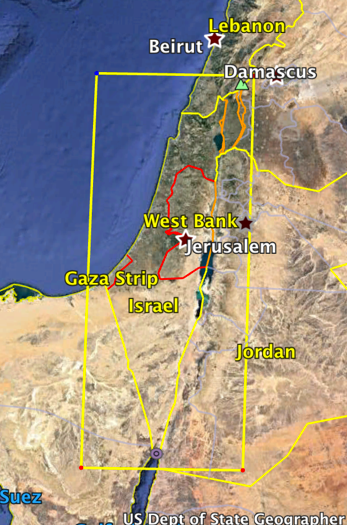}
\caption{The geo-fencing of identified coordinates. 
Coordinates that are outside of a predefined geographical area are removed.}
\label{fig:geofencing}
\end{figure}

\begin{figure}[tb]
\centering
\includegraphics[scale=0.35]{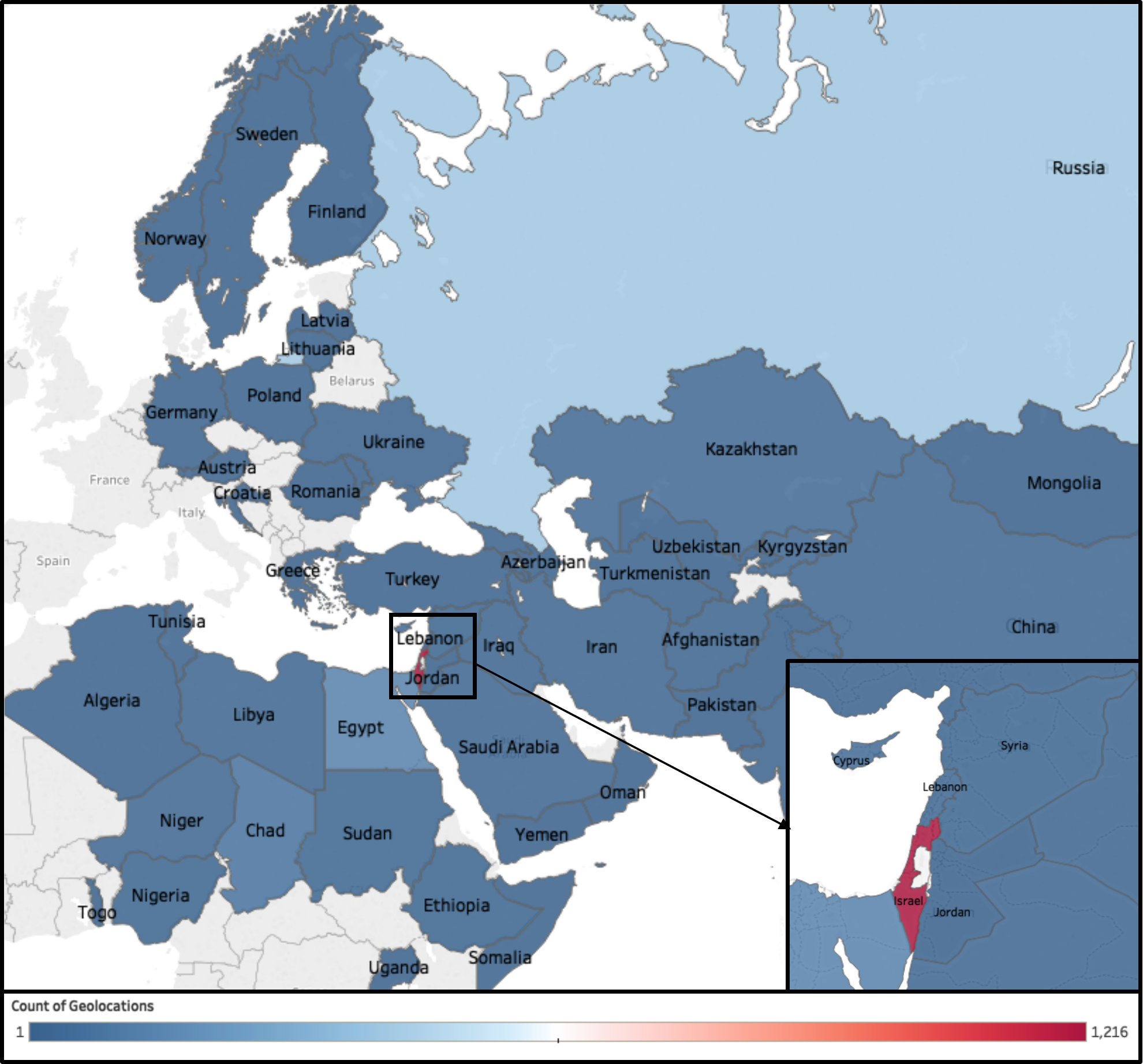}
\caption{Worldwide distribution of randomly selected geo-locations found in the network traffic for all of the experiment participants.}
\label{fig:WorldGeofencing}
\end{figure}

Each result is assigned a timestamp based on the packet capture time.
This regular expression may retrieve irrelevant results of simple float numbers which have no geographic meaning and can appear within the network traffic (e.g., the location of an object on the screen).
Therefore, in the next step we apply the following heuristics in order to filter out irrelevant results:

\begin{itemize}
\item \textbf{outgoing traffic filter.} extracting geographic coordinates from outgoing traffic (incoming traffic may contain geographic data that is not relevant to the real location of the user such as recommendations of POIs and weather forecasts);
\item \textbf{latitude/longitude pair filter.} extracting only pairs of valid geographic coordinates (i.e., those that were not already ruled out by the  previous filters) that were assigned with the same timestamp (they can appear in different IP packets due to fragmentation); this is because a coordinate is represented by two values indicating the latitude and longitude of the location.

\item \textbf{geo-fencing filter.} filtering out geographic coordinate that are outside a predefined geo-fence (e.g., the geographical boundaries of a given country or city).
In our case, all users were located within the geographical boundaries of Israel during the data collection period, and therefore we filtered out all geographic coordinates that are not within this area (as illustrated in Figure ~\ref{fig:geofencing}).
Note that in order to apply the geo-fencing filter in a general and practical one can perform reverse geo-coding on the leaked location data and identify the geographical areas that are most likely to be relevant to a user or a group of users.
We demonstrat this approach in Figure \ref{fig:WorldGeofencing},
in which we performed reverse geo-coding on randomly selected samples from the leaked location data (1\%).
As can be noticed, most of the randomly selected sample are located within the boundaries of Israel.
\end{itemize}

\subsection{Analysis and results}
In order to evaluate the amount of location data leakage, we have to determine the accuracy and correctness of the geographic coordinates detected within the network traffic.
In our experiment we could compare the geographical coordinates detected within the network traffic with the location data that was sampled by the agent application (installed on the participants' mobile phones).
By analyzing the data collected by the agent application, we observed that the location was sampled only 70\% of the time (see Figure \ref{fig:rate} for the distribution of the location sampling rate).
Possible reasons for this are that the device was off, the agent was shut down, or the location service was disabled. \\
Following the above observation, we define the \textit{active time} of a given user as the number of hours at which the agent observed at least one location sample.
Using the location samples observed by the agent, we validated the geographic coordinates that were detected within the network traffic.
Specifically, a location that observed in network traffic was classified as 'true' location only if (1) the timestamp of the detected coordinate was close enough (within 10 minutes) to a location sampled by the agent application, and (2) the measured distance between the two coordinates was below a predefined threshold of 250 meters.
If just the timestamp of the detected coordinate was close enough (within 10 minutes) to a location sampled by the agent application, we labeled the detected coordinate as 'false'; otherwise, it was labeled as 'unknown.' \\

\textbf{Volume of leaked location data.} A total of approximately 474K geolocations (complying with the standard API location regex) were identified within the network traffic of all of the monitored mobile devices.
After applying the geo-fencing filter, approximately 347K geolocations remained. 
 
\begin{figure}[tb]
\centering
\includegraphics[scale=0.4]{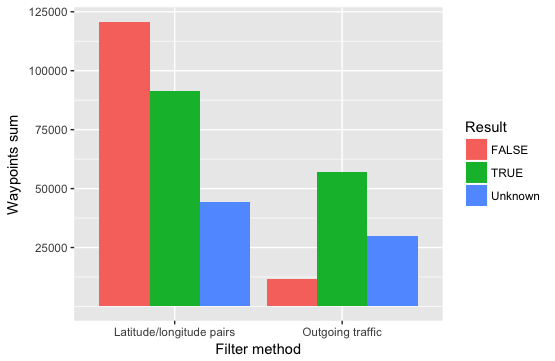}
\caption{The classification of geolocations detected within the network traffic after applying the latitude/longitude pair (left column) and after applying both the latitude/longitude pair and the outgoing traffic filters (right column).
Each column presents the distribution of labels ('true,' 'false,' and 'unknown') of the remaining geo-locations according to the labeling process described above.}
\label{fig:filtering}
\end{figure}

Figure ~\ref{fig:filtering} presents the classification of geo-locations detected within the network traffic after applying the latitude/longitude pair filter (left column) and after applying both the latitude/longitude pair filter and the outgoing traffic filter (right column).
Each column presents the distribution of labels ('true,' 'false,' and 'unknown') of the remaining geo-locations according to the labeling process described above.
In total, after applying the latitude/longitude pair filter, ~257K geolocations remained; 36\% of them were labeled as 'true,' 47\% as 'false,' and the rest could not be labeled.
After also applying the outgoing traffic filter, ~100K geo-locations remained; 58\% of them were labeled as 'true,' 11\% as 'false,' and the rest could not be labeled.

These results support our hypothesis that incoming traffic is unlikely to contain relevant geolocations of the mobile device.
We can also see that after applying all three filters, 85\% of the geolocations that could be labeled (either as 'true' or 'false') indicated the true location of the mobile device.
We can assume that the same rate also exists for the geolocations that could not be labeled (i.e., 'unknown'). \\
 
\textbf{Rate of data leakage.} By analyzing the validated geo-locations (i.e., labeled as 'true') of the 71 users, we could see that the mobile devices of about 90\% of them were leaking location traces.
The \textit{rate of data leakage} of a give user (device) is calculated by dividing the user's \textit{active time} by the number of validated leaked locations.
We partitioned the calculated leakage rate into the following groups: 'high', 'medium', 'low' and 'no-leakage' as presented in Table \ref{table:leakagerate}.

\color{black}
As can be seen in Table \ref{table:leakagerate}, 55\% of the devices were leaking location data at a medium (once every one to six hours) and high (once every one hour) rate.

\begin{table}[tb]
\centering
\begin{tabular}{|c|c|c|c|} 
\hline
Group & Leakage rate & Number of devices & Percentage\\ 
\hline
High & under 1hr & 20 & 28\%\\ 
Medium & 1-6hrs & 19 & 27\%\\
Low & 6+ hrs & 23 & 32\%\\
No leakage & $\infty$ & 9 & 13\%\\ 
\hline
\end{tabular}
\caption{The leakage rate of different mobile devices.}
\label{table:leakagerate}
\end{table}

\subsubsection{Leakage coverage.} 
\begin{figure}[tb]
\centering
\includegraphics[scale=0.4]{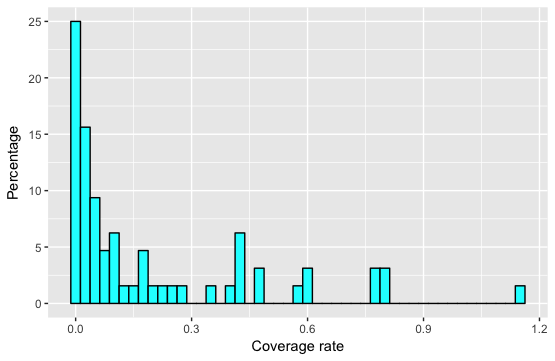}
\caption{The coverage rate measure 
is defined as the total number of exposed hours of network traffic divided by the total number of hours of agent application data.
}
\label{fig:coverage}
\end{figure}

We define an exposed hour as an hour within the collected data (network or agent) in which at least two valid location leaks were detected.\\
In order to analyze the coverage over time of relevant (validated) leaked location data, we define the \textit{coverage rate measure} as the total number of exposed hours of network traffic divided by the total number of hours of agent data (i.e., excluding hours in which the agent was not active):
\[ CoverageRate = \frac{\# of Exposed Traffic Hours}{\# of Agent Data Hours} \]
We assume that a high coverage rate will result in a high exposure and discovery rates of users' important places.
Figure ~\ref{fig:coverage} presents the distribution of the coverage rates of the mobile devices in the collected dataset. 
As can be seen, for almost 70\% of the users the coverage rate is below 0.2.

\subsubsection{Leakage inconsistency.} 
\begin{figure}[tb]
\centering
\includegraphics[scale=0.7]{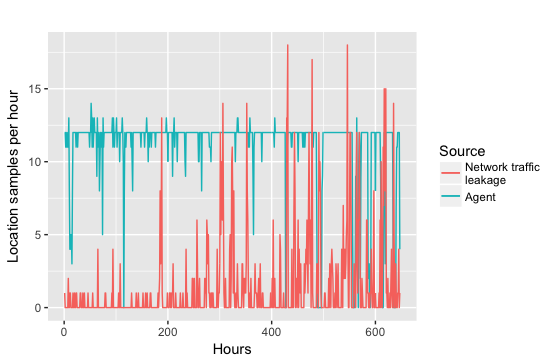}
\caption{An example of a single user, comparing the location leakage rate obtained by the agent installed on the devices and the network traffic.}
\label{fig:leakovertime}
\end{figure}

While Table \ref{table:leakagerate} and Figure \ref{fig:coverage} present the overall average leakage rate of location data, our manual exploration of the data showed that the leaked data exhibits inconsistent, non-uniform, and bursty behavior.
As an example, Figure \ref{fig:leakovertime} depicts the number of location samples per hour of a single user observed by the agent application (green line) and within the network traffic (red line).
It can be seen that while the agent's sample rate is relatively stable (around 12 samples per hour) excluding minor changes (phone shutdown or agent crash), the leaked location data within the network traffic is unstable, ranging from only a few or no leaks to a high rate of leakage.

Thus, in order to analyze and understand the inconsistency in the amount of leaked location data, we computed the relative standard deviation measure for each mobile device by dividing the standard deviation of 'leaks per hour' by the average number of 'leaks per hour.'

\begin{figure}[tb]
\centering
\includegraphics[scale=0.4]{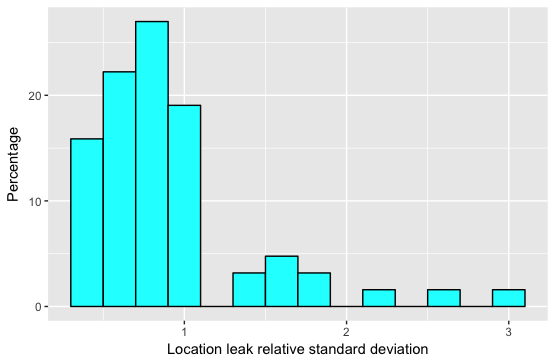}
\caption{The leakage rate variability as indicated by the leak relative standard deviation measure.}
\label{fig:relativestdev}
\end{figure}

Figure \ref{fig:relativestdev} depicts the distribution of the values of the relative standard deviation measure of the mobile devices;
a value of zero (0) indicates a constant leakage rate (i.e., zero variability).

\section{\label{sec:infferring}Inferring Points of Interest from Leaked Location Traces}
We are also interested in understanding how we can infer meaningful insights from the geolocations (coordinates) that were detected as leaks within the mobile devices' network traffic.
Specifically, we are interested in identifying a user's points of interest (POIs) and differentiating them from transit or noise data \cite{ashbrook_using_2003}.

As mentioned in the related work section, the most common a approach for identifying stay points (or POIs) is by applying clustering algorithms that are not usually bound to a predetermined number of clusters (e.g., k-means), and clustering stay points by spatial or spatio-temporal parameters. 
In this research we opted to use three different algorithms: incremental clustering \cite{kang_extracting_2005}, DBSCAN \cite{umair_discovering_2014}, and ST-DBSCAN \cite{birant_st-dbscan:_2007}. 
These algorithms usually have some assumptions on the data.
Specifically, it assumed that the data is arriving at a constant rate which was not true in our case.
Therefore, we made the following modifications to the algorithms.
First, we included the notion of time by calculating the time between samples and defined a bound on that time interval.
Second, since we processed long time frames of data, we added a phase which applies backtracking and use previously created clusters (i.e., repeating POIs) in order to increase the confidence in the identified POI.

Another approach uses semantic data to determine when a user is at an important place; for example, a location trace at a famous landmark will identify it as an important place for that user \cite{alvares_model_2007}.
This approach is not relevant in our case, because the points of interest are not known in advance; however, we use semantic data from reverse geocoding to eliminate transit geolocations (e.g., highways). 
One of the main challenges when applying clustering algorithms in fully unsupervised data is determining the parameters that will yield the best results.
This is even more challenging when the dataset is not consistent over time and for all users, as shown in Section \ref{sec:extracting}.
Based on a manual analysis of the clustering algorithm results, in our evaluation we set the thresholds for the incremental clustering algorithms at 500 meters and 30 minutes, and the value of the minPts parameter for the DBSCAN and STDBSCAN clustering to be at five pts. 



Note that we did not have any information about the users' real (confirmed) POIs in order to understand the nature and validity of the POIs identified in the network traffic leaked locations.
Therefore, as a benchmark (and ground truth) we used the POIs identified by applying the incremental clustering algorithm on the Android agent	location data (denoted as Incremental-agent).
Because the location traces collect by the mobile agent application indicate the true location of the user with high accuracy, and since POIs clustering methods have been proven in previous work, we found this benchmark sufficient for our purposes.
The POIs identified by applying the different clustering algorithms on the network traffic's leaked locations (denoted as Incremental-traffic, DBSCAN-traffic, and STDBSCAN-traffic) were compared with the agent-based POIs, namely the Incremental-agent.
We calculated the total amount of time spent at each user POI and weighed the POI significance by its part from the user total amount of time spent in all POIs. \\


\textbf{POI detection rate.} A total of 1,053 POIs (over all users) were identified by the Incremental-agent method.
For each traffic-based method we calculated: (1) the total number of POIs identified; (2) the true positive measure (number of POIs that were also detected by the Incremental-agent method); (3) the precision (the true positive value divided by the total number of POIs identified); and (4) the recall (the true positive value divided by the number of POIs identified by the benchmark method, i.e., Incremental-agent).
The results are presented in Table \ref{table:detectedPOIs}.
As can be seen, the recall, which represents the POI discovery rate, is approximately 20\% for all methods; the Incremental-traffic method yields the best results, compared to the other methods, with slightly lower recall but much higher precision.

\begin{table}[tb]
\centering
\begin{tabular}{|l|p{1.5cm}|p{1.5cm}|p{1.5cm}|} 
\hline
&Incremental-traffic&DBSCAN-traffic&STDBSCAN-traffic\\ [0.5ex]
\hline
Total         & 282   & 470   & 339   \\ 
True positive & 205   & 213   & 148    \\
Precision     & 0.73 & 0.45 & 0.43 \\
Recall        & 0.20 & 0.20 & 0.14 \\ [1ex] 
\hline
\end{tabular}
\caption{Recall and precision measures of the three methods (Incremental-traffic, DBSCAN-traffic, STDBSCAN-traffic), when considering the Incremental-agent as the ground truth of the users' POIs.}
\label{table:detectedPOIs}
\end{table}

In a real-life scenario, the adversary will not be able to label the extracted geolocations as 'true,' 'false,' or 'unknown' (as described in Section \ref{sec:extracting}).
Therefore, in Table \ref{table:detectedPOIs} we applied the clustering algorithm on \textit{all} the geolocations. \\
In Table \ref{table:Trueonly} we present the optimal results that the adversary can achieve when the POI identification process is only applied on the users' 'true' geolocations.
In this case, it can be seen that the Incremental-traffic clustering method achieved a similar recall of 20\%, however the precision improved dramatically to 95\%.

\begin{table}[tb]
\centering
\begin{tabular}{|l|p{2.5cm}|p{2.5cm}|} 
\hline
& Incremental- traffic (all) & Incremental-traffic ('true' only)\\
\hline
Precision     & 0.73 & 0.94  \\
Recall        & 0.20 & 0.20  \\
\hline
\end{tabular}
\caption{The recall and precision of the Incremental-traffic method when applied on all identified geolocations and when applied only on the confirmed ('true') geolocations.}
\label{table:Trueonly}
\end{table}

Previous work on location data analysis showed that prior semantic information (e.g., landmarks, shopping centers, roads, etc.) can be used in order to determine if a location trace is a user's POI or transit location \cite{alvares_towards_2007}.
Thus, in order to further improve the POI inference process, we used prior semantic information in order better determine the real POIs.
Specifically, we used Google's reverse geo-coding API in order to remove geo-location clusters (i.e., POIs) that are located on highways.

Table \ref{table:detectedPOIssemantic} presents the results of detected POIs (true positive, precision, and recall) when using this semantic information.
As can be seen, using semantic information to eliminate irrelevant location clusters can improve the precision with no effect on the recall.
This can be explained by the fact that due to the lower and inconsistent location leakage within the network traffic, the irrelevant location clusters (i.e., highways) are poorly reflected within the network traffic but better captured by the agent. \\

\begin{table}[tb]
\centering
\begin{tabular}{|l|p{1.5cm}|p{1.5cm}|p{1.5cm}|} 
\hline
&Incremental-traffic&DBSCAN-traffic&STDBSCAN-traffic\\ [0.5ex]
\hline
Total         & 263   & 374   & 264   \\ 
True positive & 193   & 201   & 141    \\
Precision     & 0.73 & 0.54 & 0.53 \\
Recall        & 0.20 & 0.20 & 0.14 \\ [1ex] 
\hline
\end{tabular}
\caption{The recall and precision measures of the three methods (Incremental-traffic, DBSCAN-traffic, STDBSCAN-traffic), when considering the Incremental-agent as the ground truth of the users' POIs and when using semantic information.}
\label{table:detectedPOIssemantic}
\end{table}

\textbf{The importance of the 25\% identified POIs.}
The number of identified POIs alone does not necessarily provide a good estimation for the exposure rate of users' whereabouts.
For example, let's assume a user with ten different significant locations (POIs).
If a user spends 50\% of his/her time at home and is at work 35\% of the time, by determining the user's home and work locations, we are able to identify the locations at which the user spends 85\% of his/her time (although we identified only 20\% of the user's POIs).

Thus, in order to estimate the significance of the identified locations (POIs), we computed a weighted measure for the POI detection rate as follows.
For each POI detected we computed the relative time spent by the user at that location (i.e., the total time that the user was at the POI divided by the total time the user spent at all POIs).
The weighted measure of the POIs was computed from the baseline Incremental-agent method.
Then, the POI discovery rate measure was computed by using the weights computed for each identified POI.


\noindent The results presented in Table \ref{table:exposure} show a high weighted POI discovery ratio for the medium and high leakage rates, and total of 61\% weighted POI's exposure rate. \\

\begin{table}[tb]
\centering
\begin{tabular}{|l|p{2cm}|p{2cm}|} 
\hline
Leakage rate & POI discovery ratio & Weighted POI discovery ratio \\ [0.5ex] 
\hline
High rate (under 1hr) & 48\% & 81\% \\ 
Medium rate (1-6hr) & 26\% & 67\% \\
Low rate (6+hr) & 8\% & 37\% \\
No leakage & 0 & 0 \\

\hline
Total & 27\% & 61\% \\ [1ex] 
\hline  
\end{tabular}
\caption{The POI and weighted POI discovery ratios of the Incremental-traffic method. 
The POI discovery rate is the number of places found in the network data divided by the number of places found by the benchmark method (Incremental-agent).
The weighted POI discovery rate is the amount of time spent at the identified POIs out of the total user time.}
\label{table:exposure}
\end{table}
 
\textbf{Inferring the potential exposure rate of POIs from network traffic leakage measures.}
In a real-life scenario an attacker would not have a benchmark to relate to and inferring a user's POI exposure rate would be based on captured data alone.
By deriving a regression model (Table \ref{table:linreg}) we can see that the user weighted POI exposure ratio parameter has a high correlation with the leakage rate and coverage rate measures and no significant correlation with the relative standard deviation.


\begin{table}[tb]
\centering 
 \label{} 
\begin{tabular}{@{\extracolsep{5pt}}lc} 
\\[-1.8ex]\hline 
\hline \\[-1.8ex] 
 & \multicolumn{1}{c}{\textit{Dependent variable:}} \\ 
\cline{2-2} 
\\[-1.8ex] & Weighted POI's exposure rate \\ 
\hline \\[-1.8ex] 
 Coverage & 0.559$^{***}$ \\ 
  & (0.172) \\ 
  & \\ 
 Leak rate & $-$0.0000013$^{**}$ \\ 
  & (0.00000) \\ 
  & \\ 
 Relative standard deviation & $-$0.048 \\ 
  & (0.078) \\ 
  & \\ 
 Constant & 0.597$^{***}$ \\ 
  & (0.106) \\ 
  & \\ 
\hline \\[-1.8ex] 
R$^{2}$ & 0.315 \\ 
Adjusted R$^{2}$ & 0.283 \\ 
Residual Std. Error & 0.331 (df = 65) \\ 
F Statistic & 9.955$^{***}$ (df = 3; 65) \\ 
\hline 
\hline \\[-1.8ex] 
\textit{Note:}  & \multicolumn{1}{r}{$^{*}$p$<$0.1; $^{**}$p$<$0.05; $^{***}$p$<$0.01} \\ 
\end{tabular} 
\caption{The linear regression model for estimating POIs' exposure rate from network traffic measures.}
\label{table:linreg}
\end{table} 





\section{\label{sec:discussion}Discussion}
In this section we discuss two interesting questions that arise from our research: (1) how to identify which are the leaking apps, and, (2) how can users reduce the location leakage privacy risk.

\subsection{Identifying leaking applications and services}
The goal of our analysis in this section is to determine which applications are responsible for leaking the location data and whether the leakage occurs as a result of intentional misuse or a benign application sending location data in plaintext. 
One approach for obtaining such information is real-time \textit{on device} monitoring  of the installed applications' outgoing traffic.
However, as mentioned in Section \ref{sec:datacollection}, due to security concerns such operations require super-user privileges which necessitate rooting the user's device, and therefore, such an approach was not an option in our experiment.

Deriving the information about the leaking applications from the network traffic captured is also challenging for three main reasons: (1) the network traffic captured does not provide an explicit indication of the sending application/service, (2) because of the growing use of cloud services and content delivery networks (CDNs), many destination IPs are hosted by services such as AWS, Akamai or Google, and (3) location data can be sent to advertisement and intelligence service domains by components embedded in multiple Android applications.

Given this, we opted to analyze the destination host names observed within the HTTP traffic.
We focused specifically on outgoing traffic containing location leaks.
By extracting the host names from the HTTP requests that contained the leaked location data we identified 112 different services.
Then, we analyzed the host names using public security services (VirusTotal), search engines (Google and Whois) and security reports. 
Based on the analysis results, we were able to classify each host name by its reported usage (e.g., weather forecast, navigation, location analytics, advertisement service) and whether it appears to be a legitimate or unwanted/suspicious service.
Services with clear/reasonable location usage and no reported security issues were classified as 'benign'; the rest of the services were classified as 'suspicious'.

\begin{figure}[tb]
\centering
\includegraphics[scale=0.32]{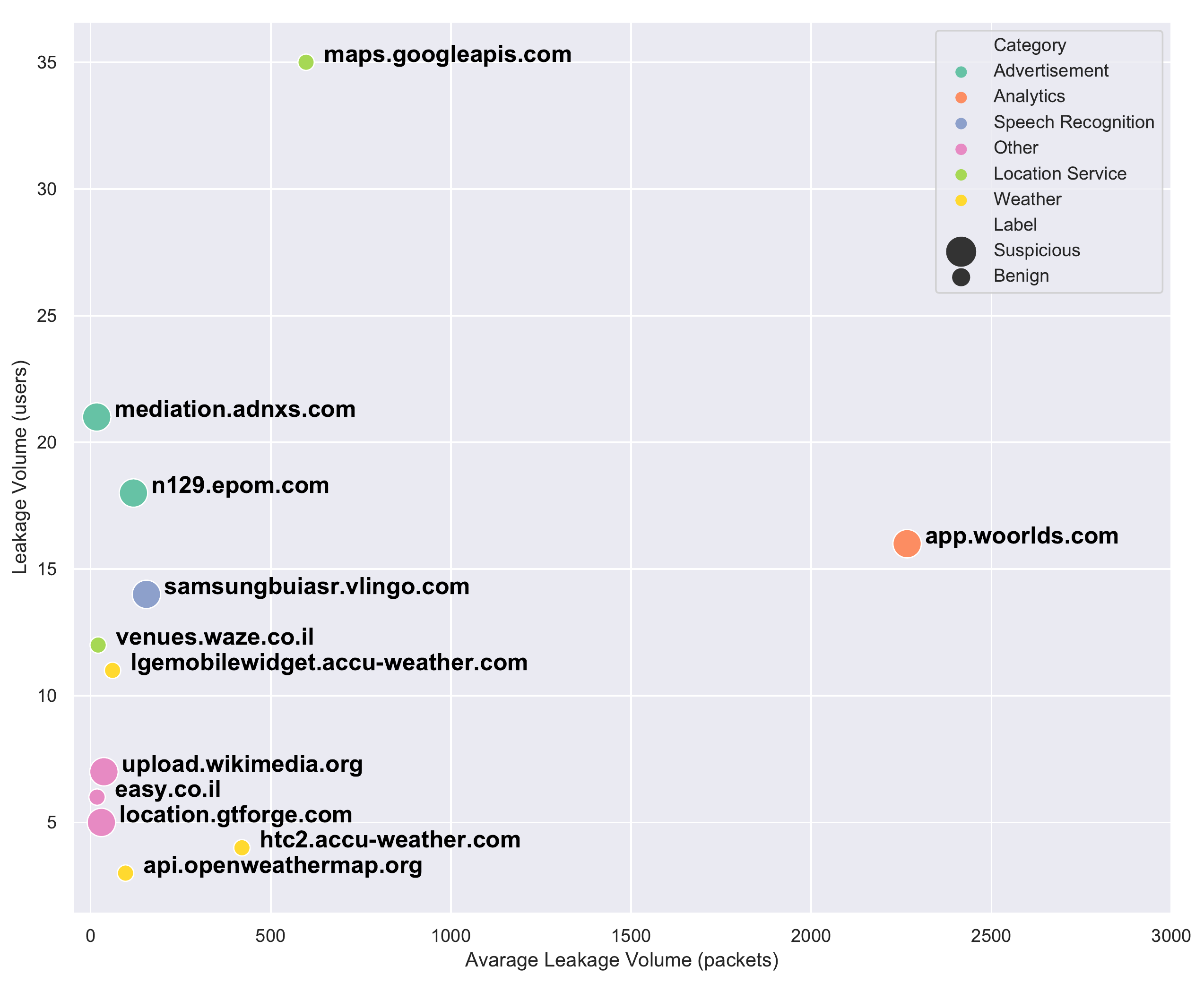}
\caption{Presenting the top 12 host names classified by their category (color) and benign/suspicious (size of circle)
The \textit{x}-axis represents the average number of detected leakage events, and the \textit{y}-axis represents the number of users involved in the experiments sending location data to that host name.}
\label{fig:servicesmapping}
\end{figure}

Figure \ref{fig:servicesmapping} presents the top 12 host names classified by their category (color) and level of suspiciousness (size of circle).
Each host name is placed on the graph according to the average number of detected leakage events (\textit{x}-axis) and the number of users involved in the experiments sending location data to that host name (\textit{y}-axis).

Some of the suspicious domain names include samsungbuiasr.vlingo.com which was previously published as a Samsung pre-installed speech recognition application named Vlingo.
This application was found to be leaking sensitive information.
Another example includes the domains, n129.epom.com and mediation.adnxs.com, which are reported to provide personalized advertisements.
An additional significant suspicious domain is app.woorlds.com, which was reported to be a location analytics service.
Interestingly, our analysis concludes that the Google Maps JavaScript API  (maps.googleapis.com), which lets Android application developers customize maps with user locations, is also responsible for sending location data in plaintext.
This is particularly noteworthy since Google recommends that application developers use the secured Maps JavaScript API (which operates over HTTPS) whenever possible.\footnote{\url{https://developers.google.com/maps/documentation/javascript/tutorial}}
Nonetheless, our analysis shows that in practice, developers also use the unsecured Map JavaScript API (which operates over HTTP).
Overall, based on the analyzed data, we found that the set of unwanted services is responsible for more than 60\% of location leakage events.

Another interesting observation from our analysis is that although the number of location data leakage events (\textit{x}-axis in Figure \ref{fig:servicesmapping}) for each individual host name is not high, based on the analysis in Section \ref{sec:infferring} we were still able identify the users' significant POIs from the data.
We attribute that finding to the fact that there are multiple applications installed on each individual smartphone, which together leak a sufficient amount of information that can be analyzed in order to infer the POIs.
%
%

Although the identity of the leaking application is not explicitly indicated in the network traffic, we performed further analysis in an attempt to link the identified host names with the applications installed on the users' mobile devices.
In order to do so, we first extracted (using our Android agent) the set of all applications that are installed on the mobile devices of the users which require both location and network permissions.
Next, we computed a modified \textit{tf-idf} measure for each pair consisting of an application and host name.
\textit{tf-idf} is a well-known measure in the field of text categorization, which is often used as a weighting factor in information retrieval and text mining \cite{sebastiani2002machine}.
The acronym \textit{tf-idf} is short for term frequency-inverse document frequency. 
It is a numerical statistic intended to reflect how important a term (i.e., word) is to a document in a collection or corpus.
The \textit{tf-idf} value increases proportionally to the number of times a term appears in
the document, but it is offset by the frequency of the term in the corpus, which helps to adjust for the fact that some terms appear more frequently in general. 
In our case, a document is a host name, and a term is an application.
The term frequency (denoted by $TF$) of an application (denoted by $a$) with respect to a given host name (denoted by $h$) is calculated as follows:

\[ TF_h(a) = \frac{|U^a_h|}{|U_h|} \]

\noindent where $U_h$ represents the set of users for which we identify a location leakage (from their devices to $h$), and $U^a_h$ represents the subset of users from $U_h$ in which application $a$ was installed on the devices.  

The inverse document frequency (denoted by $IDF$) of an application is calculated as follows:

\[ IDF(a) = -log_{10}\bigg(\frac{|U^a|}{|U|}\bigg)\]

\noindent where $U^a$ represents the set of users for which application $a$ was installed on the devices; and $U$ represents the set of all users.

Given the above, the \textit{tf-idf} of an application with respect to a given host name is calculated as follows:

\[ TFIDF_h(a) = TF_h(a) * \max\big(1,IDF(a)\big) \]

\noindent The reason for limiting the inverse document frequency ($IDF$) value to one is to prevent rare applications from achieving a very high \textit{tf-idf} score and consequently be mistakenly linked with the host name. 
In addition, we applied min-max normalization to the \textit{tf-idf} scores of applications in order to keep them within the range of zero and one.
A high $TF$ value for an application $a$ with respect to a host $h$ indicates that $a$ was frequently observed in devices that transmit location data in plaintext to $h$.
On the other hand, a high IDF value for $a$ indicates that $a$ was not frequently observed on the devices in general.
Thus, a high \textit{tf-idf} score for application $a$ with respect to host $h$ may indicate that $a$ is related to the location leakage to $h$.

\begin{figure}[tb]
\centering
\includegraphics[scale=0.28]{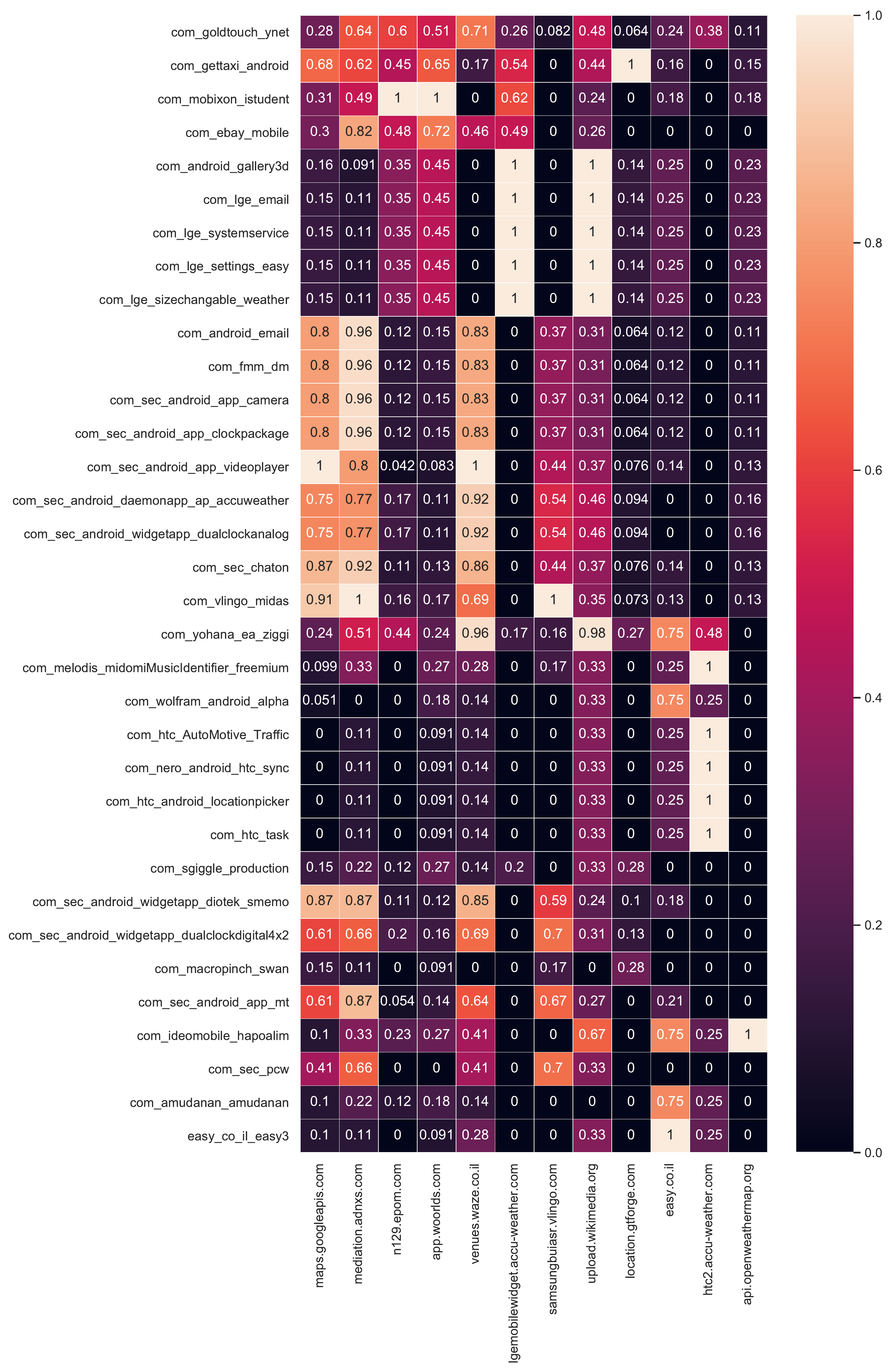}
\caption{Presenting the \textit{tf-idf} values are presented for each host name (\textit{x}-axis) and application (\textit{y}-axis).}
\label{fig:hitmap}
\end{figure}

The results of this analysis are presented in Figure \ref{fig:hitmap}, where the \textit{tf-idf} values are shown for each host name (\textit{x}-axis) and application (\textit{y}-axis).
Based on these results we can classify the applications into two categories.
The first category includes applications that send the location data to their own hosting service. 
In this category we can find multiple HTC, LG, and Samsung pre-installed applications found to be related to their own hosting services (htc2.accu-weather.com, lgemobilewidget.accu-weather.com, and samsungbuiasr.vlingo.com), as well as the GetTaxi (com.gettaxi.android) and Easy (easy.co.il.easy3) applications which were found to be related to their hosting services (location.gtforge.com and easy.co.il respectively). \\
The second category includes applications that send (usually via integrated "software plug-ins") location data to third party services such as advertisement APIs (n129.epom.com) or analytical services (app.woorlds).
In this category we identified a popular student application named com.mobixon.istudent.
We also identified applications that send location data to Google Maps services (for presenting objects on maps), some of which potentially use the Google Maps JavaScript API in an unsecured manner (the HTTP protocol instead of the HTTPS protocol).

\subsection{Mitigation strategies: reducing the privacy risk}
In order to reduce the privacy risk of location leakage and POI inference presented in our paper, the following countermeasures are suggested and could be further investigated and developed in future work.
\begin{itemize}
\item{\textbf{Awareness.}} The most basic approach is increasing the awareness of mobile device users to such risks and providing them with the tools and means to reduce the risk ~\cite{kraus2015analyzing}.
For example, reducing the risk can be achieved by installing only trusted applications (from trusted sources), monitoring and limiting sensitive permissions such as location and Internet access, and disabling location service on the device while not in use and turning it on only on demand.

\item{\textbf{OS policy and tools.}} The Android OS provides built-in standard security solutions features such as isolation, encryption, memory management, and user-granted permissions for phone resources and sensors.
Once permission has been granted by the user, PII handling is done solely by best practice recommendation \cite{Security}.
This may include enforcing encryption on applications that send sensitive information such as location over the Internet or applying a tool such as PrivacyGuard \cite{song2015privacyguard}.
In addition, future versions of the Android OS could consider improving privacy by enabling user decisions regarding background operation involving sensitive data.

\item{\textbf{Monitoring.}} Monitoring network traffic by third party security providers in order to detect PII leaks was proposed by \cite{ren_recon:_2016} and is a useful approach for identifying applications that misuse users' private information.

\item{\textbf{PII obfuscation.}} An application installed on the mobile device that monitors the location sampling or location leakage over the network traffic can intelligently inject spoofed locations that can make it difficult for a threat actor to infer the true POIs of the user.
\end{itemize}

Since the experimental setup of this research was complex and resource and time consuming, we opted leave the design, development and evaluation of the defense approaches, especially applying PII obfuscation as an adversarial learning approach on the clustering algorithms~\cite{biggio2014poisoning}, to future research.

\section{\label{sec:futurework}Conclusions and Future Work}
In this paper, we showed the extent of location leakage from mobile devices and presented a systematic process of extracting location traces from raw network traffic.
We analyzed the results of three different geo-clustering methods applied over the inconsistent network traffic data and showed that existing algorithms yield good results, even with inconsistent location data; this makes any user location privacy vulnerable even with a low rate of location leakage. 
Our work enables location exposure assessment by monitoring network traffic and reveals that even relatively low location leakage rates (e.g., six hours) and coverage of only 20\% can expose more than 70\% of the weighted POIs.
In future work we plan to further automate the POI identification process by automatically setting up the clustering algorithm parameters; evaluate the extent to which the leaked data can be used for predicting the users future location; and develop a mitigation approach based on adversarial learning techniques, which will be based on injecting a minimal number of false locations in clear text to the network traffic, in order to deceive the clustering algorithms (i.e., PII obfuscation).  

\bibliographystyle{IEEEtran}  
\bibliography{locbib} 

\begin{thebibliography}{10}
\providecommand{\url}[1]{#1}
\csname url@samestyle\endcsname
\providecommand{\newblock}{\relax}
\providecommand{\bibinfo}[2]{#2}
\providecommand{\BIBentrySTDinterwordspacing}{\spaceskip=0pt\relax}
\providecommand{\BIBentryALTinterwordstretchfactor}{4}
\providecommand{\BIBentryALTinterwordspacing}{\spaceskip=\fontdimen2\font plus
\BIBentryALTinterwordstretchfactor\fontdimen3\font minus
  \fontdimen4\font\relax}
\providecommand{\BIBforeignlanguage}[2]{{%
\expandafter\ifx\csname l@#1\endcsname\relax
\typeout{** WARNING: IEEEtran.bst: No hyphenation pattern has been}%
\typeout{** loaded for the language `#1'. Using the pattern for}%
\typeout{** the default language instead.}%
\else
\language=\csname l@#1\endcsname
\fi
#2}}
\providecommand{\BIBdecl}{\relax}
\BIBdecl

\bibitem{Nguyen:2013}
L.~T. Nguyen, Y.~Tian, S.~Cho, W.~Kwak, S.~Parab, Y.~S. Kim, P.~Tague, and
  J.~Zhang, ``Unlocin: Unauthorized location inference on smartphones without
  being caught,'' 06 2013.

\bibitem{hazan2017dynamic}
I.~Hazan and A.~Shabtai, ``Dynamic radius and confidence prediction in
  grid-based location prediction algorithms,'' \emph{Pervasive and Mobile
  Computing}, vol.~42, pp. 265--284, 2017.

\bibitem{michalevsky_powerspy:_2015}
Y.~Michalevsky, A.~Schulman, G.~A. Veerapandian, D.~Boneh, and G.~Nakibly,
  ``{PowerSpy}: {Location} {Tracking} {Using} {Mobile} {Device} {Power}
  {Analysis}.'' in \emph{{USENIX} {Security} {Symposium}}, 2015, pp. 785--800.

\bibitem{arigela_mobile_2013}
L.~Arigela, P.~Veerendra, S.~Anvesh, and K.~Satya, ``Mobile {Phone} {Tracking}
  \& {Positioning} {Techniques},'' \emph{International Journal of Innovative
  Research in Science, Engineering and Technology}, vol.~2, no.~4, 2013.

\bibitem{kang_extracting_2005}
J.~H. Kang, W.~Welbourne, B.~Stewart, and G.~Borriello, ``Extracting places
  from traces of locations,'' \emph{ACM SIGMOBILE Mobile Computing and
  Communications Review}, vol.~9, no.~3, pp. 58--68, 2005.

\bibitem{umair_discovering_2014}
M.~Umair, W.~S. Kim, B.~C. Choi, and S.~Y. Jung, ``Discovering personal places
  from location traces,'' in \emph{Advanced {Communication} {Technology}
  ({ICACT}), 2014 16th {International} {Conference} on}.\hskip 1em plus 0.5em
  minus 0.4em\relax IEEE, 2014, pp. 709--713.

\bibitem{ren_recon:_2016}
\BIBentryALTinterwordspacing
J.~Ren, A.~Rao, M.~Lindorfer, A.~Legout, and D.~Choffnes,
  ``\BIBforeignlanguage{en}{{ReCon}: {Revealing} and {Controlling} {PII}
  {Leaks} in {Mobile} {Network} {Traffic}}.''\hskip 1em plus 0.5em minus
  0.4em\relax ACM Press, 2016, pp. 361--374. [Online]. Available:
  \url{http://dl.acm.org/citation.cfm?doid=2906388.2906392}
\BIBentrySTDinterwordspacing

\bibitem{regulation2016general}
P.~Regulation, ``General data protection regulation,'' \emph{Official Journal
  of the European Union}, vol.~59, pp. 1--88, 2016.

\bibitem{taylor_longitudinal_2016}
V.~F. Taylor and I.~Martinovic, ``A longitudinal study of app permission usage
  across the google play store,'' \emph{CoRR, abs/1606.01708}, 2016.

\bibitem{grolman_transfer_2018}
E.~Grolman, A.~Finkelstein, R.~Puzis, A.~Shabtai, G.~Celniker, Z.~Katzir, and
  L.~Rosenfeld, ``Transfer {Learning} for {User} {Action} {Identication} in
  {Mobile} {Apps} via {Encrypted} {Trafc} {Analysis},'' \emph{IEEE Intelligent
  Systems}, vol.~PP, no.~99, pp. 1--1, 2018.

\bibitem{hodges_android_2014}
\BIBentryALTinterwordspacing
D.~Hodges, ``\BIBforeignlanguage{en}{Android apps and privacy risks : what
  attackers can learn by sniffing mobile device traffic},'' 2014. [Online].
  Available:
  \url{https://ora.ox.ac.uk/objects/uuid:17c44695-402c-4275-8dab-468966d8fe0b}
\BIBentrySTDinterwordspacing

\bibitem{ren_bug_2018}
\BIBentryALTinterwordspacing
J.~Ren, M.~Lindorfer, D.~J. Dubois, A.~Rao, D.~Choffnes, and
  N.~Vallina-Rodriguez, ``\BIBforeignlanguage{en}{Bug {Fixes}, {Improvements},
  ... and {Privacy} {Leaks} - {A} {Longitudinal} {Study} of {PII} {Leaks}
  {Across} {Android} {App} {Versions}}.''\hskip 1em plus 0.5em minus
  0.4em\relax Internet Society, 2018. [Online]. Available:
  \url{https://www.ndss-symposium.org/wp-content/uploads/sites/25/2018/02/ndss2018_05B-2_Ren_paper.pdf}
\BIBentrySTDinterwordspacing

\bibitem{Ester:1996:DAD:3001460.3001507}
\BIBentryALTinterwordspacing
M.~Ester, H.-P. Kriegel, J.~Sander, and X.~Xu, ``A density-based algorithm for
  discovering clusters a density-based algorithm for discovering clusters in
  large spatial databases with noise,'' in \emph{Proceedings of the Second
  International Conference on Knowledge Discovery and Data Mining}, ser.
  KDD'96.\hskip 1em plus 0.5em minus 0.4em\relax AAAI Press, 1996, pp.
  226--231. [Online]. Available:
  \url{http://dl.acm.org/citation.cfm?id=3001460.3001507}
\BIBentrySTDinterwordspacing

\bibitem{birant_st-dbscan:_2007}
\BIBentryALTinterwordspacing
D.~Birant and A.~Kut, ``\BIBforeignlanguage{en}{{ST}-{DBSCAN}: {An} algorithm
  for clustering spatial–temporal data},'' \emph{\BIBforeignlanguage{en}{Data
  \& Knowledge Engineering}}, vol.~60, no.~1, pp. 208--221, Jan. 2007.
  [Online]. Available:
  \url{http://linkinghub.elsevier.com/retrieve/pii/S0169023X06000218}
\BIBentrySTDinterwordspacing

\bibitem{montoliu_discovering_2010}
R.~Montoliu and D.~Gatica-Perez, ``Discovering human places of interest from
  multimodal mobile phone data,'' in \emph{Proceedings of the 9th
  {International} {Conference} on {Mobile} and {Ubiquitous}
  {Multimedia}}.\hskip 1em plus 0.5em minus 0.4em\relax ACM, 2010, p.~12.

\bibitem{alvares_model_2007}
L.~O. Alvares, V.~Bogorny, B.~Kuijpers, J.~A.~F. de~Macedo, B.~Moelans, and
  A.~Vaisman, ``A model for enriching trajectories with semantic geographical
  information,'' in \emph{Proceedings of the 15th annual {ACM} international
  symposium on {Advances} in geographic information systems}.\hskip 1em plus
  0.5em minus 0.4em\relax ACM, 2007, p.~22.

\bibitem{khan2018empirical}
M.~T. Khan, J.~DeBlasio, G.~M. Voelker, A.~C. Snoeren, C.~Kanich, and
  N.~Vallina-Rodriguez, ``An empirical analysis of the commercial vpn
  ecosystem,'' in \emph{Proceedings of the Internet Measurement Conference
  2018}.\hskip 1em plus 0.5em minus 0.4em\relax ACM, 2018, pp. 443--456.

\bibitem{mani2018understanding}
A.~Mani, T.~Wilson-Brown, R.~Jansen, A.~Johnson, and M.~Sherr, ``Understanding
  tor usage with privacy-preserving measurement,'' in \emph{Proceedings of the
  Internet Measurement Conference 2018}.\hskip 1em plus 0.5em minus 0.4em\relax
  ACM, 2018, pp. 175--187.

\bibitem{ISO6709}
``{Standard representation of geographic point location by coordinates}, volume
  = {2008}, address = {Geneva, CH}, institution = {International Organization
  for Standardization},'' Standard, Jul. 2008.

\bibitem{Location}
``{Location-Android developer documentation},''
  \url{https://developer.android.com/reference/android/location/Location.html/},
  2018, [Online; accessed 12-April-2018].

\bibitem{ashbrook_using_2003}
D.~Ashbrook and T.~Starner, ``Using {GPS} to learn significant locations and
  predict movement across multiple users,'' \emph{Personal and Ubiquitous
  computing}, vol.~7, no.~5, pp. 275--286, 2003.

\bibitem{alvares_towards_2007}
L.~O. Alvares, V.~Bogorny, B.~Kuijpers, B.~Moelans, J.~A. Fern, E.~D. Macedo,
  and A.~T. Palma, ``Towards semantic trajectory knowledge discovery,''
  \emph{Data Mining and Knowledge Discovery}, vol.~12, 2007.

\bibitem{sebastiani2002machine}
F.~Sebastiani, ``Machine learning in automated text categorization,'' \emph{ACM
  computing surveys (CSUR)}, vol.~34, no.~1, pp. 1--47, 2002.

\bibitem{kraus2015analyzing}
L.~Kraus, T.~Fiebig, V.~Miruchna, S.~M{\"o}ller, and A.~Shabtai, ``Analyzing
  end-users’ knowledge and feelings surrounding smartphone security and
  privacy,'' \emph{S\&P. IEEE}, 2015.

\bibitem{Security}
``{Security Tips-Android developer documentation},''
  \url{https://developer.android.com/training/articles/security-tips.html/},
  2018, [Online; accessed 01-April-2018].

\bibitem{song2015privacyguard}
Y.~Song and U.~Hengartner, ``Privacyguard: A vpn-based platform to detect
  information leakage on android devices,'' in \emph{Proceedings of the 5th
  Annual ACM CCS Workshop on Security and Privacy in Smartphones and Mobile
  Devices}.\hskip 1em plus 0.5em minus 0.4em\relax ACM, 2015, pp. 15--26.

\bibitem{biggio2014poisoning}
B.~Biggio, K.~Rieck, D.~Ariu, C.~Wressnegger, I.~Corona, G.~Giacinto, and
  F.~Roli, ``Poisoning behavioral malware clustering,'' in \emph{Proceedings of
  the 2014 workshop on artificial intelligent and security workshop}.\hskip 1em
  plus 0.5em minus 0.4em\relax ACM, 2014, pp. 27--36.

\end{thebibliography}
\end{document}